\documentclass[12pt]{article}
\usepackage[dvipdfmx]{graphicx}
\usepackage{amssymb}
\usepackage{amsmath}
\usepackage{bm}
\usepackage{braket}
\usepackage{color}
\usepackage[colorlinks=true,linkcolor=blue,citecolor=blue]{hyperref}

\begin{document}

\begin{titlepage}

\begin{flushright}
ICRR-Report-602-2011-19\\
IPMU~11-0213
\end{flushright}

\vskip 1.35cm

\begin{center}

{\large 
{\bf Gravitational waves from smooth hybrid new inflation}
}

\vskip 1.2cm

Masahiro Kawasaki$^{a,b}$, 
Ken'ichi Saikawa$^c$
and
Naoyuki Takeda$^a$ \\

\vskip 0.4cm

{ \it$^a$Institute for Cosmic Ray Research,
University of Tokyo, 5-1-5 Kashiwa-no-ha, Kashiwa City, Chiba, 277-8582, Japan}\\
{\it $^b$Kavli Institute for the Physics and Mathematics of the Universe, 
Todai Institutes for Advanced Study, 
The University of Tokyo, 5-1-5 Kashiwa-no-ha, Kashiwa City, Chiba 277-8568, Japan}\\
{\it $^c$Department of Physics, Tokyo Institute of Technology, 2-12-1 Ookayama, 
Meguro-ku, Tokyo 152-8511, Japan}\\

\date{\today}

\begin{abstract}
We calculate the production of the gravitational waves from a double inflation model with lattice simulations. 
Between the two inflationary stages, gravitational waves with a characteristic frequency are produced by fluctuations of the scalar fields enhanced through parametric resonance. 
The wavelength of the produced gravitational waves gets extra redshift during the second inflationary stage and it can be in the observable range for the direct gravitational wave detectors. 
It is found that there is a possibility for the produced gravitational waves to be detected  in the planned experiments.\\
\end{abstract}

\end{center}
\end{titlepage}

\tableofcontents


\section{Introduction}
\label{sec:intro}


The general relativity proposed by Einstein predicts the existence of 
the gravitational waves, which are the distortions of space-time geometry 
propagating through space as waves. 
Though it is indirect, there is strong evidence for the existence of the gravitational waves which 
is provided by the observation of the binary pulsar $1913+16$ \cite{Hulse:1974eb}. 
Yet up to date no direct detection has been made. 
In the world, there are lots of ongoing and planned experiments to detect the 
gravitational waves directly. 
Laser Interferometer Gravitational-Wave Observer~(LIGO)~\cite{LIGO} 
is working on the ground in USA.
In Europe, Virgo~\cite{Virgo} commenced operations in 2007 and 
Einstein Telescope~(ET)~\cite{ET} is planned. 
In Japan, KAGRA~\cite{KAGRA} is now under construction.
It is also planned to construct the space-borne interferometers such as 
Laser Interferometer Space Antenna~(LISA)~\cite{LISA},~
Deci-hertz Interferometer Gravitational Wave Observatory~(DECIGO)~
\cite{DECIGO1,Seto:2001qf,Kawamura:2011zz}, 
and Big Bang Observer~(BBO)~\cite{Crowder:2005nr}. 

The direct detection experiments of the gravitational waves are expected to provide rich information 
about various astrophysical phenomena, but one of the ultimate goals of the experiments is to probe 
the early history of the universe. 
It is presumed that in the very early universe the vacuum energy dominates the 
universe and causes quasi-exponential expansion of the universe 
called inflation. 
Inflation explains 
the homogeneity and flatness of the universe. 
Planck satellite~\cite{Ade:2013zuv}
has measured anisotropies of the cosmic microwave background~(CMB) 
and successfully determined the cosmological parameters with high precision, which 
supports the prediction of the  inflationary theories.  
However, it is impossible to prove directly the era before photons decouple from electrons, 
if we use the result of the observation of cosmic microwave background only. 
In order to study the early universe before photon decoupling, 
we use the gravitational wave 
since the gravitational wave interacts with other particles very 
weakly and preserves the information about the early universe. 

In this paper, we discuss the gravitational wave production from inflation. 
There are several primordial origins of the gravitational waves: quantum 
fluctuations of the metric~\cite{Lyth:2009zz}, 
preheating~
\cite{Khlebnikov:1997di,Easther:2006gt,
 GarciaBellido:2007af,Dufaux:2008dn}, 
domain walls~\cite{Hiramatsu:2010yz,Kawasaki:2011vv}, cosmic strings~
\cite{Berezinsky:2000vn,Kawasaki:2010yi,Damour:2000wa,Damour:2004kw}, 
and so on. 
In this paper, we focus on the preheating as the origin of gravitational waves. 
After the end of the inflation, the inflaton rapidly oscillates around the true 
vacuum, interacting with other scalar fields. At this epoch, fluctuations of the
scalar fields enhanced vastly~
\cite{GarciaBellido:1997wm,Felder:2000hj}, 
which produce abundant amount of gravitational waves. 
However, in general, the wavelength of the gravitational waves produced from 
reheating is so short~\cite{Dufaux:2008dn} that we cannot detect the gravitational waves 
even in the future gravitational wave experiments with high sensitivity.  
In this work, we show that the double inflation model, 
which was originally proposed in~
\cite{Randall:1995dj,GarciaBellido:1996qt,
Izawa:1997df}
to solve the initial value problem 
for the new inflation,  
produces gravitational waves with the wavelength long enough to be proved with the 
planned detectors. 

In the double inflation, there are two inflationary 
stages. 
In this paper, we consider the smooth 
hybrid new inflation model~\cite{Yamaguchi:2004tn}. 
At the first stage, smooth hybrid inflation~
\cite{Lazarides:1995vr,Jeannerot:2000sv} occurs, 
followed by the second stage at 
which new inflation occurs. 
In the intermediate stage, the inflaton and waterfall fields oscillate around the minimum of the 
potential in the same way as the reheating. 
Since in the smooth hybrid inflation, 
the gauge symmetry of the waterfall fields is 
broken even during the inflation, no harmful topological defects are produced. 
In the previous study~\cite{Kawasaki:2006zv}, it was found that in this model 
fluctuations of the scalar fields 
are enhanced by the parametric resonance and then primordial black holes are 
formed at a characteristic scale. 
In this paper, we pay attention to the production of the gravitational waves from 
these fluctuations. 
Since the parametric resonance is a nonlinear phenomenon, 
we calculate the production of the gravitational waves with lattice simulations.

The organization of this paper is as follows.
We review the double inflation model in Sec.~\ref{sec:Smooth hybrid inflation}.
In Sec.~\ref{sec:gw_production}, we calculate the amount of the produced gravitational waves by using analytical consideration and lattice simulations.
We estimate the present density of the gravitational waves and the peak frequency in Sec.~\ref{sec:present_density}. 
Finally, Sec.~\ref{sec:conclusion} is devoted to the conclusion.


\section{Smooth hybrid new inflation}
\label{sec:Smooth hybrid inflation}


In this section, we review the smooth hybrid new inflation model 
in supergravity proposed in \cite{Yamaguchi:2004tn}. 
This model has two inflationary stages. 
At first, the smooth hybrid inflation \cite{Lazarides:1995vr}, 
whose gauge symmetry is already broken at the beginning 
of the inflation differently from the hybrid 
inflation model based on supergravity~
\cite{Linde:1997sj}, 
takes place. 
After the first inflationary stage, 
the new inflation~\cite{Izawa:1996dv} follows. 
We assume that the $e$-fold number of the new inflation is smaller than $~60$. 
Thus, the density fluctuations on large scales are produced 
during the smooth hybrid inflation. 
In this section, we set the reduced planck mass $ M_{\mathrm{pl}} \simeq 2.4\times10^{18}{\mathrm{GeV}}$ equal to unity unless otherwise stated.

\subsection{Smooth hybrid inflation}
First, we discuss the smooth hybrid inflation model. 
This model has three superfields. 
One is the inflaton field $ S$ and the others are waterfall fields ${\Psi}$ and $\bar{\Psi}$. 
This model is based on $ U(1)_R$ symmetry. 
The superpotential is given by
\begin{equation}
 W_H = S\left(-{\mu}^2+\frac{(\bar{\Psi}\Psi)^2}{M^2}\right),
\end{equation}
where $ S$ has $R$ charge $ 2$ and $(\bar{\Psi}\Psi)$ has $R$ charge $0$ 
and $M$ is a cut-off scale. 
The $R$-invariant K\"{a}hler potential is given by
\begin{equation}
 K_H = |S|^2 + |\Psi|^2 + |\bar{\Psi}|^2.
\end{equation}
We also introduce the $U(1)$ gauge symmetry under which the waterfall fields are 
transformed as
$ \Psi\rightarrow e^{i\delta}\Psi$ 
and $\bar{\Psi}\rightarrow e^{-i\delta}\bar{\Psi}$, and $Z_2$ symmetry 
under which $\left(\Psi\bar{\Psi}\right)$ has a unit charge. 
Using the phase rotation and D-term flat condition, we can bring the 
complex scalar fields $\Psi$ and $\bar{\Psi}$ on the real axis 
as $\psi\equiv2{\rm Re}\Psi=2{\rm Re}\bar{\Psi}$.

From the superpotential and the K\"{a}hler potential, we can write down the scalar potential. 
Using the $R$ symmetry, we can bring the complex scalar field $ S$ on the 
real axis, 
$ {\sigma} \equiv \sqrt{2}ReS$. 
Neglecting higher-order terms, we can write down the scalar potential for $\sigma<1$
\begin{equation}
 V_H \simeq 
\left(-{ \mu}^2+\frac{{\psi}^4}{16M^2}\right)^2\left(1+\frac{{\psi}^2}{2} 
	+\frac{{\sigma}^4}{8}\right)+
\frac{{\sigma}^2{\psi}^6}{16M^4}-
{\mu}^2\frac{{\sigma}^2{\psi}^4}{4M^2},
\end{equation}
where 
$ \sigma$ is the inflaton and 
$ \psi$ is the waterfall field. 
During inflation, $ \psi$ is located at the local minimum of the potential, 
\begin{eqnarray}
\left\{
\begin{array}{l}
{\psi}_{\mathrm {min}} \simeq \sqrt{\frac{4}{3}}\frac{{\mu}M}{{\sigma}}
	\hspace{5ex}{\mathrm {for}} \hspace{1ex}{\sigma}\gg \sqrt{{\mu}M},\\
{\psi}_{\mathrm {min}} \simeq 2\sqrt{{ \mu}M}
	\hspace{5ex}{\mathrm {for}} \hspace{1ex}{\sigma}\ll \sqrt{{\mu}M}.
\end{array}
\right.
\end{eqnarray}
Since the gauge symmetry of the waterfall fields $\Psi$ and $\bar{\Psi}$ is 
already broken during inflation, no topological defects are formed in this model.
When $\sigma$ is larger than $\sqrt{\mu M}$, the effective potential is given by 
\begin{equation}
 V_H({ \sigma}) \simeq 
{\mu}^4\left[1-\frac{2}{27}\frac{{\mu}^2M^2}{{\sigma}^4}+\frac{{\sigma}^4}{8}\right].
\end{equation}
The third term reflects the effect of the supergravity. 
Since the effective potential is dominated by the false vacuum energy 
${ \mu}^4$, the Hubble parameter is given by
\begin{equation}
 H \simeq \frac{{ \mu}^2}{\sqrt{3}}.
\end{equation}

To consider the dynamics of the inflation, we differentiate the effective potential as 
\begin{equation}
\frac{{\partial} V_H}{{\partial}{ \sigma}} =
{ \mu}^4\left[\frac{8}{27}\frac{({\mu}M)^2}{{ \sigma}^5}
 + \frac{1}{2}{ \sigma}^3\right].
\end{equation}
The second term, which reflects the effect of the supergravity, dominates the inflation dynamics 
when inflaton $ \sigma$ is larger than $ {\sigma}_d$ given by
\begin{equation}
{ \sigma}_d = \left(\frac{16}{27}\right)^{1/8}\left({\mu}M\right)^{1/4}.
\end{equation}
Let us define $\sigma_{i}$ as the value of the inflaton field 
when the pivot scale $k_0=0.002[{\mathrm {Mpc}}^{-1}]$ leaves the horizon.  
The slow-roll condition is broken when $|{ \eta}| = |\frac{1}{2}\frac{V''}{V}|\sim1$ 
and inflation ends at 
$ { \sigma} \simeq { \sigma}_c$ 
where
\begin{equation}
 { \sigma}_c = \left(\frac{40}{27}\right)^{1/6}\left({ \mu}M\right)^{1/3}.
\end{equation}
Then, the number of $e$-folds $N_H$ from $\sigma_{i}$ to $\sigma_c$ is written as
\begin{equation}
N_{\mathrm{H}}
=
\int^{\sigma_{i}}_{\sigma_{c}}d\sigma
\frac{V}{V'}
=
\int^{\sigma_{i}}_{\sigma_{c}}d\sigma
\frac{1}{\frac{8}{27}\frac{\left(\mu M\right)^2}{\sigma^5}+\frac{1}{2}\sigma^3}.
\end{equation}
The amplitude of the curvature perturbation 
in comoving gauge $\cal R$ and the spectral index $n_s$ at the pivot scale are written as 
\begin{equation}
{\cal R}
 \simeq
\frac{1}{2\pi\sqrt{3}}\frac{V^{3/2}(\sigma_{i})}{|V'(\sigma_{i})|}
=
\frac{\mu^2}{
	2\pi\sqrt{3}\left[
		\frac{8}{27}\frac{\left(\mu M\right)^2}{\sigma_i^5}
		+\frac{1}{2}\sigma_i^3
	\right]
}
=
\frac{\mu^2}{
\pi\sqrt{3}\left[
	\frac{\sigma_d^8}{\sigma_i^5}	+\sigma_i^3
\right]
}
,
\end{equation}
and
\begin{equation}
n_s-1
=
-\frac{80}{27}\frac{\left(\mu M\right)^2}{\sigma_{i}^6}
+
3\sigma_{i}^2
=
-\frac{1}{5}\frac{\sigma_d^8}{\sigma_i^6}
+3\sigma_i^2
.
\end{equation}
In this model, when the value of the inflaton at the pivot scale 
$\sigma_{i}$ is larger than $\sigma_{d}$, 
the spectral index becomes blue. 
\subsection{New inflation}
Second, we describe the new inflation model~\cite{Izawa:1996dv}.
This model has one superfield ${ \Phi}$. 
${ \Phi}$ has $R$ charge $\frac{2}{5}$, and it is assumed that $ U(1)_{R}$ 
symmetry is 
dynamically broken down to a discrete $ Z_{8R}$ at a scale $ v$. 
The superpotential is given by 
\begin{equation}
{ W_{N} }= { v}^{{}2}{ \Phi} - \frac{ g}{5}{\Phi}^{5},
\end{equation}
where $g$ is a coupling constant. 
The $R$-invariant K\"{a}hler potential is given by
\begin{equation}
 K_{N} = |{ \Phi}|^{2} + \frac{C_{N}}{4}|{ \Phi}|^{4}.
\end{equation}
Here, $ C_{N}$ is a constant of order 1.

The scalar potential yields a vacuum,
\begin{equation}
 \braket{\Phi} \simeq \left(\frac{ v^{2}}{ g}\right)^{{1/4}}.
\end{equation}
At this vacuum, the potential has the negative vacuum energy as
\begin{equation}
{ V(\braket{ \Phi}) \simeq}
{ -3 e^{K_{N}}|W_{N}\left[\braket{ \Phi}\right]|^{2} \simeq}
{ -3\left(\frac{4}{5}\right)^{2}}{ v}^{ 4}\left(\frac{{ v}^{2}}{{ g}}\right)^{1/2}.
\end{equation}
We assume that the negative vacuum energy is cancelled out by 
a supersymmetry-breaking effect which gives a positive contribution 
${ \Lambda}_{\mathrm{SUSY}}^{4}$ to the vacuum energy. 
Thus, we have a relation between $ v$ and the gravitino mass $ m_{3/2}$ as 
\begin{equation}
{ m_{3/2} \simeq}
{ \frac{{ \Lambda}_{\mathrm{SUSY}}^{2}}{\sqrt{3}}=}
{\frac{4}{5}}{ v}^{ 2}\left(\frac{{ v}^{ 2}}{ g}\right)^{ 1/4}.
\end{equation}

To discuss the dynamics of the new inflation, 
we identify the inflaton field $\phi$ with the real part of the field $\Phi$ as 
$\phi \equiv \sqrt{2}Re{ \Phi}$. 
Neglecting the higher-order terms, we obtain the scalar potential of the 
new inflation as
\begin{equation}
{ V_{H}\left[ \phi\right] \simeq}
{ v}^{4} -{ \frac{C_{N}}{2}}{ v}^{ 4}{ \phi}^{ 2}
-\frac{ g}{ 2}{ v}^{ 2}{ \phi}^{ 4}
+\frac{{ g}^{ 2}}{ 16}{ \phi}^{ 8}.
\end{equation}
We differentiate the scalar potential as
\begin{equation}
{\frac{dV_{N}}{d{\phi}} = }
{ -C_{N}}{ v}^{ 2}{ \phi} 
{ -2}{ g}{ v}^{ 2}{ \phi}^{ 3}
+\frac{{ g}^{2}}{ 2}{ \phi}^{ 7}.
\end{equation}
When ${\phi}$ is smaller than ${\phi}_{d}\equiv\sqrt{C_Nv^2/(2g)}$, 
the first term dominantly controls the inflation dynamics. 
On the other hand, when $ \phi$ is larger than ${ \phi}_{d}$, 
the second term dominantly controls the inflation dynamics. 
New inflation occurs when the inflaton $ \phi$ is smaller than $ { \phi}_{c}$ at 
which slow-roll condition is broken,
$|{ \eta}|=\frac{1}{2}|\frac{V_{N}''}{V_{N}}|\sim1$, and $\phi_c$ is given by
\begin{equation}
{ {\phi}_{c} =}
\sqrt{\frac{{ v}^{ 2}}{{ 6}{ g}} {\left(1-C_{N}\right)}}.
\end{equation}
Then, the number of e-folds $ N_{\mathrm{new}}$ is 
written as 
\begin{equation}
 N_{\mathrm{new}} \simeq
\frac{1}{C_{N}}\ln\frac{{\phi}_{d}}{{ \phi}_{i}} 
+ \frac{1-4C_{N}}{2C_{N}\left(1-C_{N}\right)}.
\end{equation}
Here we have assumed that $ C_{N}\leq \frac{1}{4}$. 
\subsection{Oscillatory phase}
Here, let us consider the oscillatory phase from the end of the 
smooth hybrid inflation to the start of the new inflation. 

After the end of the smooth hybrid inflation, $ \sigma$ and $ \psi$ 
roll down toward ${ \sigma}_{\mathrm{min}} = 0$ and $ { \psi}_{\mathrm{min}} = 2\sqrt{{ \mu}M}$
where the potential has the minimum. 
Around the potential minimum, their effective masses $ m_{ \sigma}$ and $ m_{ \psi}$ are 
given by
\begin{equation}
\begin{array}{l}
 m_{ \sigma} = \sqrt{\frac{8{\mu}^{3}}{M}}\gg H,\\[0.5\intextsep]
 m _{ \psi} = \sqrt{\frac{8{\mu}^{3}}{M} + 16{ \mu}^{4}}\gg H.
\end{array}
\end{equation}
Thus, $ \sigma$ and $ \psi$ oscillate around their respective minima and 
the total energy density decreases as $ a^{-3}$, 
where $a$ is the scale factor. 
Eventually, the false vacuum energy of the new inflation $ v^{4}$ 
dominates the total energy density of the universe, and then new inflation starts. 
During this oscillatory phase, the effective masses of the inflaton and waterfall fields change periodically through their mutual couplings, which leads to the rapid amplification of the field fluctuations ${{\delta}{\sigma}}$ and ${{\delta}{\psi}}$ called the parametric resonance~\cite{Landau,Shtanov:1994ce,Kofman:1997yn}. 
In the present model, the most amplified fluctuations have wavenumber $ k_{\mathrm{peak}}$~\cite{Kawasaki:2006zv} given by 
\begin{equation}\label{eq:2:kpeak}
 k_{\mathrm{peak}} \simeq 0.35 m_{ \sigma}.
\end{equation}
We expect that these enhanced fluctuations 
produce gravitational waves at the scale $ k_{\mathrm{peak}}$.

Next, we consider the initial condition for the new inflation. 
Here, interactions between the two inflation sectors become important. 
During the smooth hybrid inflation, the relevant part of the potential 
is given by 
\begin{equation}
{ V_{\mathrm{int}} = \frac{1}{2}{ \mu}^{4}{ \phi}^{2}} + 
{ \mu}^{ 2}{ v}^{ 2}{ \sigma}{ \phi}.
\end{equation}
Thus, at the end of the smooth hybrid inflation, the value of the inflaton of the 
new inflation is given by
\begin{equation}
{{\phi}_{\mathrm{min}}=-}\frac{{ v}^{2}}{{\mu}^{2}}{\sigma}_{ c}.
\end{equation}
During the oscillatory phase, 
the potential energy of the scalar fields averaged over one oscillation time 
is the half of the total energy density of the universe, 
which leads to the effective mass of $\phi$,
\begin{equation}\label{eq:nmass}
 m_{ \phi}^2 
 \simeq \frac{3H^2}{2}.
\end{equation}
Taking into account $ \dot{H} =-\frac{3}{2}H^{2}$, and using eq.~(\ref{eq:nmass}), 
one can find that the amplitude of $ \phi$ decreases as $ a^{-3/4}$. 
Thus, the initial condition for the new inflation is estimated as
\begin{equation}
{ { \phi}_{\mathrm{ini}}\simeq-}\frac{{ v}^{3}}{{\mu}^{3}}{\sigma}_{ c}.
\end{equation}

Taking into account this oscillatory phase, and assuming 
the ordinary thermal history after the reheating, 
we estimate the total number of $e$-folds $ N_{\mathrm{tot}}$ as
\begin{equation}\label{eq:2:ntot}
 N_{\mathrm{tot}}
=  N_{H} + N_{\mathrm{new}}
=  57.6 + \frac{2}{3}\ln\mu + \frac{1}{3}\ln \left(\frac{T_{\mathrm{R}}}{10^9\mathrm{GeV}}\right),
\end{equation}
where we set the reheating temperature as $T_{\mathrm{R}}$.

\subsection{Parameter search for $\sigma_{i}$}
In this subsection, we search for the allowed region of $\sigma_{i}$ at 
each $n_s$.
At the pivot scale $k_0$, the spectral index~$n_s$ and 
the curvature perturbation~$\cal R$ are written in terms of 
$\sigma_{i}$, the energy scale of the smooth hybrid inflation~$\mu$ 
and the cut-off scale~$M$. 
Then, using the Planck constraint ${\cal R}\simeq4.7\times10^{-5}$~
\cite{Ade:2013zuv},
 we can write down $\mu$
and $M$ in terms of $n_s$ and $\sigma_{i}$ as 
\begin{equation}\label{eq:mu}
\mu 
= 7.2\times10^{-3}~
\sqrt{
8~\sigma_{i}^3-\left(n_s-1\right)\sigma_{i}
},
\end{equation}
\begin{equation}\label{eq:M}
M
=
81~\sigma_{i}^3
\sqrt{\frac{
3~\sigma_{i}^2-\left(n_s-1\right)
}{
8~\sigma_{i}^3-\left(n_s-1\right)\sigma_{i}
}}.
\end{equation}
The value of the inflaton $\sigma_{c}$, at which the smooth hybrid inflation ends, 
is also written in terms of $\sigma_{i}$ and $n_s$ as
\begin{equation}
\sigma_{c}
=
0.89~\sigma_{i}
\left[3~\sigma_{i}^2-\left(n_s-1\right)\right]^{1/6}.
\end{equation}
The $e$-folding number of the smooth hybrid inflation $N_{\mathrm{H}}$ and that of the 
new inflation $N_{\mathrm{new}}$ are written as 
\begin{equation}
\begin{split}
N_{\mathrm{H}}
&=
\int^{\sigma_{i}}_{\sigma_c}d\sigma
\frac{\mu^4}{\mu^4\left[
\frac{8}{27}\frac{\left(\mu M\right)^2}{\sigma^5}+\frac{\sigma^3}{2}
\right]}\\
&=
\int^{\sigma_{i}}_{
  0.89~\sigma_{i}
  \left[3~\sigma_{i}^2-\left(n_s-1\right)\right]^{1/6}
}d\sigma
\frac{1}{
  \frac{1}{10}\frac{
    \sigma_{i}^6\left(3\sigma_{i}^2-\left(n_s-1\right)\right)
  }{\sigma^5}
  + \frac{\sigma^3}{2}
},
\end{split}
\end{equation}
\begin{equation}\label{eq:Nn}
\begin{split}
N_{\mathrm{new}}
&=
N_{\mathrm{tot}}-N_{\mathrm{H}}\\
&=
54.3+\frac{1}{3}
\ln \left[
8\sigma_{i}^3-\left(n_s-1\right)\sigma_{i}
\right]\\[0.5\intextsep]
&\hspace{5ex}
-
\int^{\sigma_{i}}_{
  0.89~\sigma_{i}
  \left[3~\sigma_{i}^2-\left(n_s-1\right)\right]^{1/6}
}d\sigma
\frac{1}{
  \frac{1}{10}\frac{
    \sigma_{i}^6\left(3\sigma_{i}^2-\left(n_s-1\right)\right)
  }{\sigma^5}
  + \frac{\sigma^3}{2}
},
\end{split}
\end{equation}
where we have used (\ref{eq:2:ntot}) and (\ref{eq:mu}) for $N_{\mathrm{tot}}$ and $\mu$.
Thus, the parameters of the smooth hybrid new inflation are written in terms of 
$\sigma_{i}$ and $n_s$. 

Since the effective potential in the supergravity is uncontrollable over 
the planck mass, 
$\sigma_{i}$ and $\mu$ must be smaller than the planck mass. 
In addition to this, $\sigma_i$ must be larger than $\sigma_c$. 
This condition guarantees that the hybrid inflation occurs. 
After the hybrid inflation, the fluctuations of the scalar fields are 
amplified through the parametric resonance, and then these fluctuations 
lead to a sharp peak at the scale corresponding to $k_{\rm peak}^{-1}$ 
in the spectrum of the curvature perturbation~\cite{Kawasaki:2006zv}. 
The observations of the large scale structures have not seen such 
enhancement of the curvature perturbations so that the scale 
corresponding to $k_{\rm peak}^{-1}$ should be sufficiently small. 
Thus, $N_{\mathrm{H}}$ must be larger than about $10$. 
Furthermore, since we assume that the new inflation occurs after the 
hybrid inflation, $N_{\mathrm{new}}$ should be larger than $0$. 
In addition to these conditions, the cut-off scale $M$ must be larger 
than $\mu$. 
These conditions are summarized as
\begin{eqnarray}\label{eq:condition}
\left\{
\begin{array}{ll}
(\mathrm{i}) &
  \sigma_{c}<\sigma_{i}<1\\[0.5\intextsep]
(\mathrm{i}\hspace{-.1em}\mathrm{i}) &
  \mu < 1\\[0.5\intextsep]
(\mathrm{i}\hspace{-.1em}\mathrm{i}\hspace{-.1em}\mathrm{i}) &
  \mu < M
  \\[0.5\intextsep]
(\mathrm{i}\hspace{-.1em}\mathrm{v}) &
  N_{\mathrm{H}}>10
  ,~
  N_{\mathrm{new}}>0
\end{array}
\right.
\end{eqnarray}
Substituting (\ref{eq:mu})$-$(\ref{eq:Nn}) into (\ref{eq:condition}), 
we can obtain the maximum and the minimum values of $\sigma_{i}$.

\begin{figure}[htbp]
\begin{center}
\begin{tabular}{c c}
\resizebox{120mm}{!}{\includegraphics{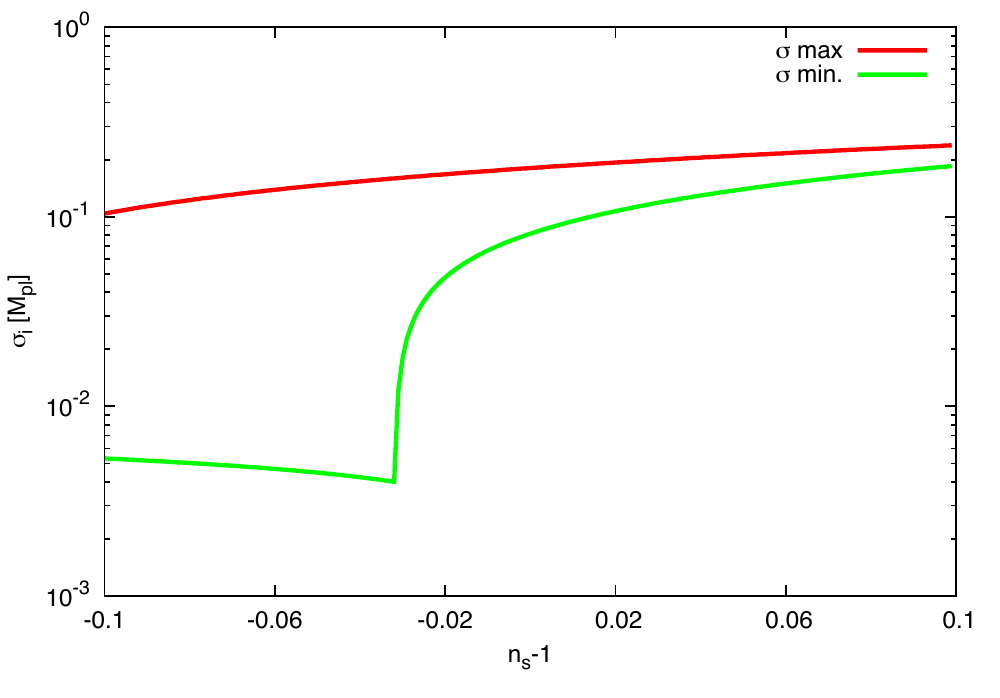}}\\
\resizebox{120mm}{!}{\includegraphics{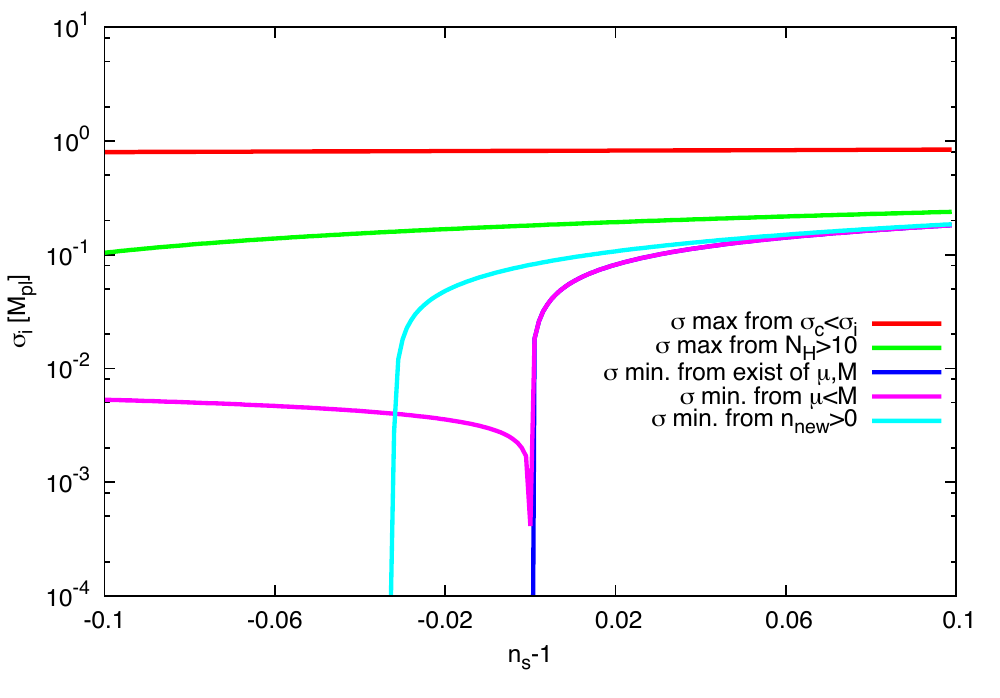}} 
\end{tabular}
\caption{
Maximum  and Minimum values of $\sigma_{i}$ at each value of 
$n_s$ obtained from the set of conditions (\ref{eq:condition}). 
The upper panel shows the maximum and minimum values of 
$\sigma_{i}$, satisfying  all conditions of (\ref{eq:condition}).
The lower panel shows each maximum and minimum value of $\sigma_{i}$, 
satisfying each condition given by (\ref{eq:condition}).
The red line is the upper limit satisfying $\sigma_c < \sigma_i$.
The green line is the upper limit satisfying $N_{\mathrm{H}} > 10$.
The purple line is the lower bound on $\sigma_i$ coming from $\mu < M$. 
The lower limit on $\sigma_i$ from $N_{\mathrm{new}} > 0$ is represented by the sky blue line.
The blue line shows the lower limit for existence of $\mu$ and $M$. 
}
\label{fig:simami}
\end{center}
\end{figure}
\begin{figure}[htbp]
\begin{center}
\begin{tabular}{c c}
\resizebox{120mm}{!}{\includegraphics{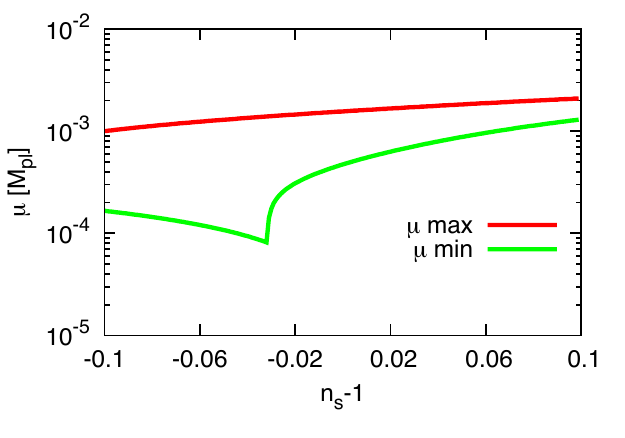}} \\
\resizebox{120mm}{!}{\includegraphics{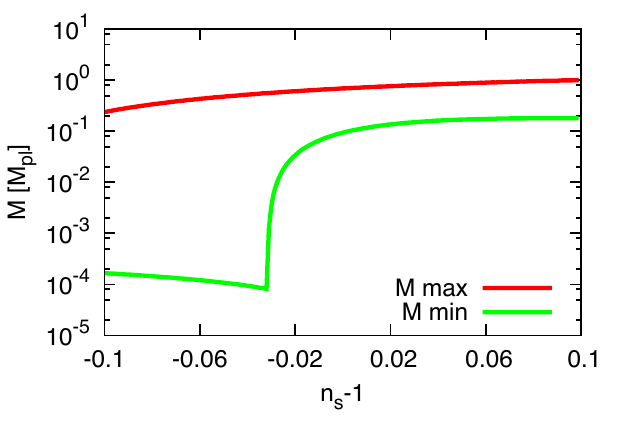}}
\end{tabular}
\caption{
The upper(lower) panel shows the maximum and minimum values of $\mu$($M$) 
at each values of $n_s$, satisfying  all conditions of 
(\ref{eq:condition}).
}
\label{fig:muM}
\end{center}
\end{figure}
Fig.~\ref{fig:simami} shows the maximum and the minimum values of 
$\sigma_{i}$ at each value of $n_s$.
Using (\ref{eq:mu}) and (\ref{eq:M}), we can get the corresponding maximum and the minimum values of $\mu$ and $M$ which are shown in~Fig.~\ref{fig:muM}.


\section{Production of the gravitational waves}
\label{sec:gw_production}


During the oscillatory phase, the fluctuations of the scalar fields grow exponentially through the parametric resonance. 
These growing fluctuations can be a source for gravitational waves. 
In this section, we calculate the amount of the produced gravitational waves. 
First, we roughly calculate the energy density of the gravitational waves. 
Second, we introduce the formalism with which 
we calculate the energy density and the spectrum of the gravitational waves 
produced from the scalar fields. 
Finally, we perform the lattice simulations and calculate the evolution of the 
fluctuations of the scalar fields. 
Using the result of the lattice simulations, we estimate the energy density and the 
spectrum of the gravitational waves.

\subsection{Order estimate}
In this subsection, we roughly calculate the ratio 
$\Omega_{\mathrm{gw},p}$ of the energy density of the gravitational waves 
to the total energy density 
based on the naive estimation given by Felder and Kofman~
\cite{Dufaux:2008dn,Felder:2006cc}. 
Here and hereafter the subscript $p$ means the epoch, at which gravitational waves are produced. 

Through the parametric resonance, the fluctuations of the scalar fields with the typical length scale 
$R_{\mathrm{peak}}\simeq k_{\mathrm{peak}}^{-1}$ are enhanced most. 
These fluctuations produce the gravitational waves. 
The enhanced fluctuations have self gravitational potential energy 
$E_{\mathrm{gw}}$. 
We can write down $E_{\mathrm{gw}}$  at $t_p$ as 
\begin{equation}
E_{\mathrm{gw}}
\sim
G\frac{
\left(\rho_{\mathrm{tot},p}R_{\mathrm{peak}}^3\right)^2
}{R_{\mathrm{peak}}}
\simeq
\rho_{\mathrm{tot},p}H_{p}^2R_{\mathrm{peak}}^5,
\end{equation}
where $\rho_{\mathrm{tot},p}$ is the total density and $H_{p}$ is the Hubble parameter. 
We approximate that the order of the energy of the gravitational waves is 
comparable to the self gravitational potential energy of the 
enhanced scalar fields $E_{\mathrm{gw}}$. 
Under this approximation,
the energy density of the gravitational waves  
$\rho_{\mathrm{gw},p}$ is estimated as 
\begin{equation}
\rho_{\mathrm{gw},p}
\sim
\frac{E_{\mathrm{gw}}}{R_{\mathrm{peak}}^3}
\sim
\rho_{\mathrm{tot},p}H_{p}^2R_{\mathrm{peak}}^2.
\end{equation}
Thus, the ratio of the energy density of the gravitational waves to the 
total energy density $\Omega_{\mathrm{gw},p}$ is given by
\begin{equation}
   \Omega_{\mathrm{gw},p}  
   \equiv \frac{\rho_{\mathrm{gw},p}}{\rho_{\mathrm{tot},p}}
   \sim \left(R_{\mathrm{{peak}}}H_{p}\right)^2.
\end{equation}
In this paper we assume that the density of the produced gravitational waves is given by the above equation with introducing a numerical constant as 
\begin{equation}\label{eq:3:Omegap0}
   \Omega_{\mathrm{gw},p}
   = \alpha\left(R_{\mathrm{{peak}}}H_{p}\right)^2, 
\end{equation}
where $\alpha$  will be determined by the numerical simulations presented in Sec.~\ref{subsec:simulation}. 
For the smooth hybrid new inflation model,  
from~(\ref{eq:2:kpeak}), $R_{\mathrm{peak}}$ is written  as
\begin{equation}
R_{\mathrm{peak}}
=
\frac{1}{0.35m_{\sigma}}=\frac{1}{0.35}\sqrt{\frac{M}{8\mu^3}}.
\end{equation}
Using this relation, we can rewrite (\ref{eq:3:Omegap0}) as 
\begin{equation}\label{eq:3:Omegap}
   \Omega_{\mathrm{gw},p}
   =0.34~\alpha\left(\frac{\mu M}{M_{\mathrm{pl}}^2}\right).
\end{equation}
\subsection{Formalism for calculating gravitational wave spectrum}
Here, we review the formalism developed by Dufaux et al.~\cite{Dufaux:2007pt} 
with which we calculate the spectrum of the gravitational waves produced 
from the fluctuations of the scalar fields. 

We consider the Friedmann-Robertson-Walker metric including the linear 
perturbation in the spatial metric ${\cal H}_{ij}$. 
\begin{equation}
ds^2
=
a^2(\tau)
\left[
  -d\tau^2 + 
  \left(
  \delta_{ij} + {\cal H}_{ij}(\tau,\vec{x})
  \right)dx^{i}dx^{j}
\right],
\end{equation}
where $\tau$ is the conformal time defined as $dt = ad\tau$. 
The amplitude of the gravitational waves $h_{ij}$ is defined as the 
transverse-traceless part of ${\cal H}_{ij}$ 
\begin{equation}
h_{ij}(\tau,\vec{x})
=
a(\tau){\cal H}_{ij}^{\mathrm{TT}}(\tau,\vec{x}).
\end{equation}
The superscript TT means the transverse-traceless part.
The evolution of $h_{ij}$ is described by the linearized Einstein 
equation as 
\begin{equation}\label{eq:3:heq}
h_{ij}''(\tau,\vec{x})-\nabla^2h_{ij}(\tau,\vec{x})
=
16\pi Ga(\tau)T_{ij}^{\mathrm{TT}}(\tau,\vec{x}).
\end{equation}
where prime denote the differential with respect to the conformal time. 
Assuming that the typical physical scale is much smaller than the horizon scale,
we neglected the terms including $a''$. 
Here, $T_{ij}^{\mathrm{TT}}(\tau,\vec{x})$ is computed by applying the projection 
operator in the momentum space as
\begin{equation}
\begin{split}
 &T_{ij}^{TT}(\tau,\vec{k})=\Lambda_{ij,kl}(\hat{k})T_{ij}(\tau,\vec{k})
 =\Lambda_{ij,kl}\left\{\partial_k\phi\partial_l\phi\right\}(\tau,\vec{k}),\\
 &\Lambda_{ij,kl}(\hat{k}) 
 = P_{ik}(\hat{k})P_{jl}(\hat{k})  -\frac{1}{2}P_{ij}(\hat{k})P_{kl}(\hat{k}),\\
 &P_{ij}(\hat{k}) = \delta_{ij}-\hat{k}_{i}\hat{k}_{j},
\end{split}
\end{equation}
where $\left\{\partial_k\phi\partial_{l}\phi\right\}(\tau,\vec{k})$~is 
the Fourier transform of 
$\partial_{k}\phi(\tau,\vec{x})\partial_{l}\phi(\tau,\vec{x})$. 

We assume that the source term is nonzero only for the time interval 
$ {\tau}_{i}\leq{\tau}\leq{\tau}_{f}$. 
In this case we can solve (\ref{eq:3:heq}) by using the 
Green's function and obtain  
\begin{equation}\label{eq:3:h}
{h_{ij}}({\tau},\vec{ k})
= { A_{ij}}(\vec{ k}){ \sin\left[k\left({ \tau}-{ \tau}_{f}\right)\right]}
 +{ B_{ij}}(\vec{ k}){ \cos\left[k\left({ \tau}-{ \tau}_{f}\right)\right]}
 \hspace{5ex}{ \mathrm{for}}\hspace{1ex}{\tau}\geq{\tau}_{ f},
\end{equation}
where
\begin{equation}
\begin{split}
&{ A_{ij}}(\vec{ k})
 ={\frac{-16\pi G}{k}\int^{{\tau}_{f}}_{{ \tau}_{i}}d{\tau'}
 	\cos\left[k\left({\tau}_{f}-{\tau}'\right)\right]a({ \tau}')
		T_{ij}^{TT}}({ \tau}',\vec{ k}),\\
&{ B_{ij}}(\vec{ k})
 ={\frac{16\pi G}{k}\int^{{\tau}_{f}}_{{ \tau}_{i}}d{\tau'}
 	\sin\left[k\left({\tau}_{f}-{\tau}'\right)\right]a({ \tau}')
		T_{ij}^{TT}}({ \tau}',\vec{ k}).
\end{split}
\end{equation}
Here, $h_{ij}(\tau,\vec{k})$~is the Fourier transform of $h_{ij}(\tau,\vec{x})$. 

The energy density of the gravitational waves is given by 
\begin{equation}
\rho_{\mathrm{gw}}(\tau)
=
\frac{1}{32\pi Ga^4(\tau)}
\braket{
h_{ij}'(\tau,\vec{x}) h_{ij}'(\tau,\vec{x})
}_V
=
\frac{1}{32\pi Ga^4(\tau)V}
\int\frac{d^3k}{\left(2\pi\right)^3}
h_{ij}'(\tau,\vec{k})h_{ij}^{\ast'}(\tau,\vec{k}),
\end{equation}
where $\langle \cdots\rangle_V$ denotes the average over the spatial volume $V$.
Taking the time average over a period of the oscillations, we obtain 
$\rho_{\mathrm{gw}}(\tau)$ as
\begin{equation}
\begin{split}
\rho_{\mathrm{gw}}(\tau)
&=
\frac{4\pi G}{a^4(\tau)V}
\int\frac{d^3k}{\left(2\pi\right)^3}
\sum_{ij}\\
&\hspace{1ex}\times
\left[
\left|
\int^{\tau_{f}}_{\tau_{i}}d\tau'
\cos\left(k\tau'\right)a\left(\tau'\right)T_{ij}^{TT}(\tau',\vec{k})
\right|^2
+
\left|
\int^{\tau_{f}}_{\tau_{i}}d\tau'
\sin\left(k\tau'\right)a\left(\tau'\right)T_{ij}^{TT}(\tau',\vec{k})
\right|^2
\right].
\end{split}
\end{equation}
Thus, the  spectrum of the gravitational waves just after the production is 
given by 
\begin{equation}\label{eq:3:OmegaL}
\Omega_{\mathrm{gw}}\left(k\right)
=
\frac{1}{\rho_{\mathrm{tot}}}
\frac{d\rho_{\mathrm{gw}}(\tau_{f},k)}{d\ln k}
=
\frac{4G^2}{3\pi Va^4(\tau_{f})H^2(\tau_{f})}
S_{k}(\tau_{f}),
\end{equation}
where
\begin{equation}\label{eq:3:sk}
\begin{split}
S_{k}(\tau_{f})
&=
k^3\int d\Omega_{k}
\sum_{ij}\\
&\times
\left[
\left|
\int^{\tau_{f}}_{\tau_{i}}d\tau'
\cos\left(k\tau'\right)a\left(\tau'\right)T_{ij}^{TT}(\tau',\vec{k})
\right|^2
+
\left|
\int^{\tau_{f}}_{\tau_{i}}d\tau'
\sin\left(k\tau'\right)a\left(\tau'\right)T_{ij}^{TT}(\tau',\vec{k})
\right|^2
\right],
\end{split}
\end{equation}
and $d\Omega_{k}=d\cos\theta d\phi$.

\subsection{Lattice simulation}
\label{subsec:simulation}

The parametric resonance is a nonlinear phenomenon. 
Thus, in order to estimate the spectrum and energy density of produced 
gravitational waves precisely, we need to perform the lattice simulations
to calculate the time evolution of the fluctuations of the scalar fields. 

We perform the lattice simulations with three dimensional comoving box taking the cosmic expansion into account, 
where the scale factor is determined by the Friedmann equation. 
In order to resolve the oscillations of the inflaton, we set the time step $dt$ as
\begin{equation}
dt = \frac{1}{10}\frac{1}{m_{\sigma}}<\frac{1}{m_{\sigma}}.
\end{equation}
We set the step number of the time as $1500$, where the ratio of the scale factor 
at the end of the simulation to that at the beginning is~$3.1$
\footnote{
If we continue the simulation for a larger time step, 
the energy density of the universe is dominated by the constant energy 
density $v^4$ and the scale factor increases exponentially with time, 
which determines the upper limit of the simulation time. 
}. 
By the end of the simulation, the production of the gravitational 
waves stops. 
The linear analysis~\cite{Kawasaki:2006zv} shows that the enhancement of the 
scalar fields occurs at the spacial scale 
$R_{\mathrm{peak}}\sim1/\left(0.35m_{\sigma}\right)$. 
Taking this into account, we set the box size of the simulation $V=L^3$ and the 
grid number $N_{\mathrm{grid}}$ as 
$L=100/m_{\sigma}$ and $N_{\mathrm{grid}}=256^3$. 
Then, $L>R_{\mathrm{peak}}$ and $L/N_{\mathrm{grid}}^{1/3}<R_{\mathrm{peak}}$ 
are satisfied even if we consider the growth of the scale factor.
Thus, we can resolve the peak scale $R_{\mathrm{peak}}$ during the simulation. 
We choose the physical parameters as follows 
\begin{equation}\label{eq:3:parameter}
{\mu} = { 1.38\times10^{-3}M_{{\rm pl}}},~
{ v} = { 3.45\times10^{-4}M_{{\rm pl}}},~
{ M} = { 0.545M_{{\rm pl}}},~
{ g} = { 2.0\times10^{-5}},~\mathrm{and}~
{ C_{N} = 0.04},
\end{equation}
with which the spectral index $n_s-1=-0.035$ is within 1 $\sigma$ 
deviation from the central value of the Planck result~\cite{Ade:2013zuv}. 
We set the initial amplitude of the fluctuations of the scalar fields by  
Rayleigh distribution~\cite{Polarski:1995jg,Khlebnikov:1996mc,
Khlebnikov:1996zt,GarciaBellido:2002aj}.
\begin{figure}[tbp]
\begin{center}
\begin{tabular}{c c}
\resizebox{145mm}{!}{\includegraphics{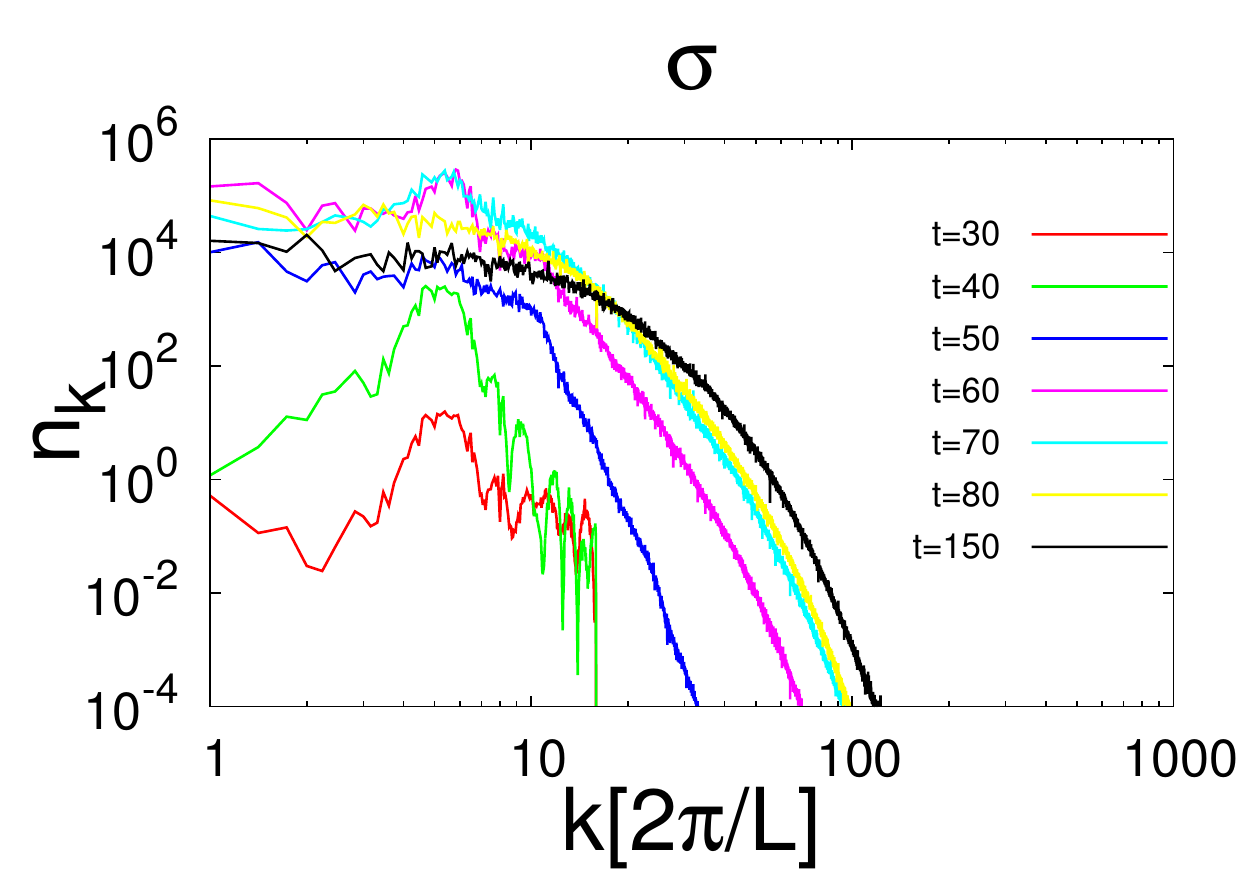}} \\
\resizebox{145mm}{!}{\includegraphics{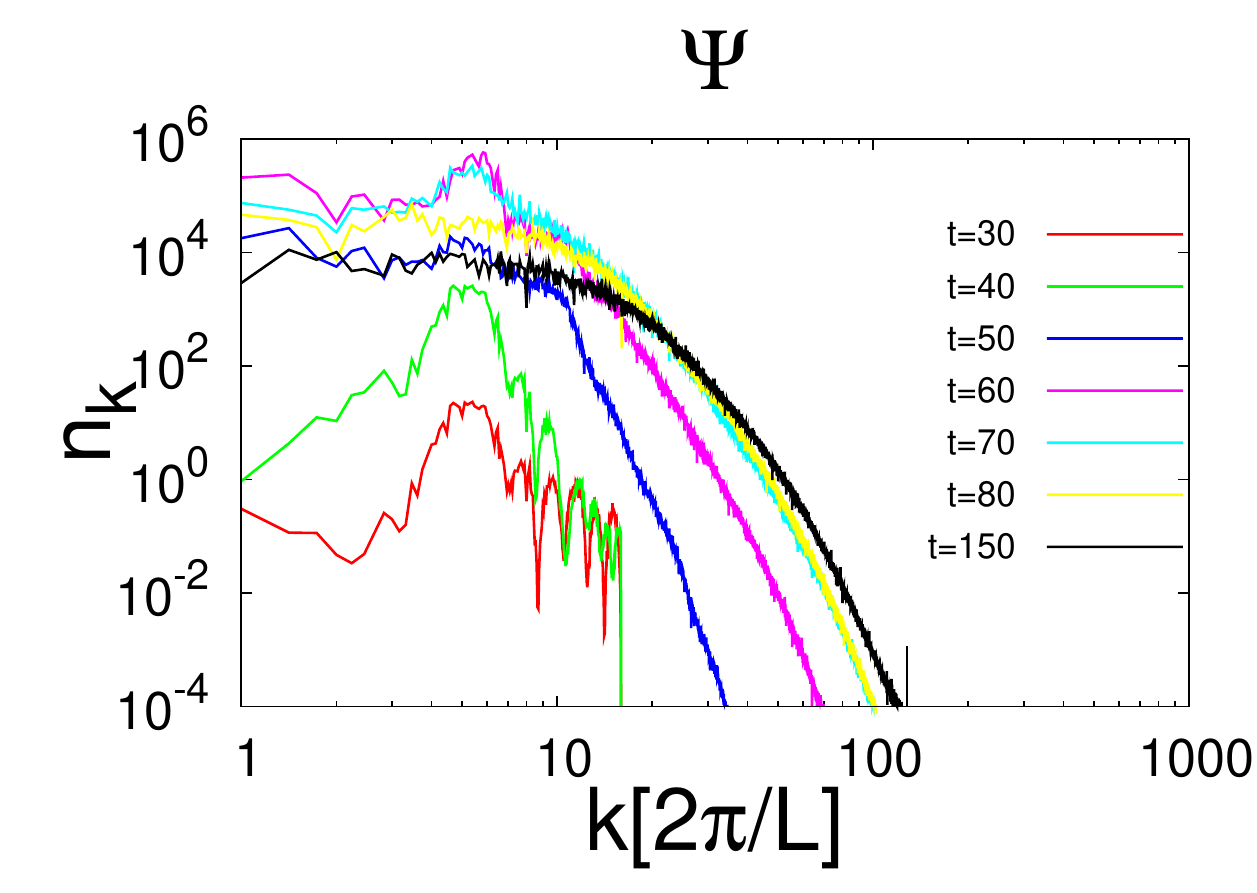}} 
\end{tabular}
\caption{
Time evolution of the occupation number $n_{f}$ from $t = 30[1/m_{\sigma}]$
to $t=80[1/m_{\sigma}]$ for every $10[1/m_{\sigma}]$ time interval and 
at final time $t=150[1/m_{\sigma}]$.
The top and bottom panels show the time evolutions of $n_f$ for $\sigma$ 
and $\phi$ respectively.}\label{fig:3:n_real}
\end{center}
\end{figure}

The inflation model we consider here has three scalar fields $\sigma,~\psi$~and
$\phi$, where
$\sigma$ is the inflaton of the hybrid inflation, 
$\psi$ is the waterfall field of the hybrid inflation and
$\phi$ is the inflaton of the new inflation. 
We perform the simulation using these three real scalar fields.\footnote{

More precisely, this model has three complex scalar fields 
$S,~\Psi$~and~$\Phi$. 
The real part of $S,~\Psi$ and $\Phi$ corresponds to 
$\sigma,~\psi$ and $\phi$. 
The imaginary part of each complex scalar fields does not affect on the dynamics of 
inflation directly. 
However, there is a possibility that the fluctuations of their imaginary parts 
are enhanced through the parametric resonance and the shape of the 
gravitational waves becomes different from that simulated only by three real 
scalar fields. 
We have performed the simulations with these three complex scalar fields, from which we confirmed that the spectrum is not altered by the fluctuations of the imaginary part.}
Using the lattice simulations, we calculate the occupation number of the 
scalar fields in momentum space $n_{f}$ as 
\begin{equation}
n_{f}
=
\omega_{k}\left(
\frac{|\dot{f}_{k}|^2}{\omega_{k}^2} + |f_{k}|^2
\right),~
\omega_{k}^2 \equiv \frac{k^2}{a^2}+m_{f}^2:~f\in\{\sigma,~\psi,~\phi\}, 
\end{equation}
where $f_{k}$ is the Fourier component of $f$.
Fig.~\ref{fig:3:n_real} shows the time evolution of $n_f$ for the scalar 
fields. 
We can see the enhancement of the scalar fields $\sigma$ and $\psi$ 
by the parametric resonance. 
\footnote{
We have checked that the fluctuations of the inflaton for the new inflation 
$\phi$ are not enhanced. This means that the inflaton of the new inflation 
does not affect the dynamics of the oscillatory phase.
}

\begin{figure}[tbp]
\begin{center}
\includegraphics[width=12cm]{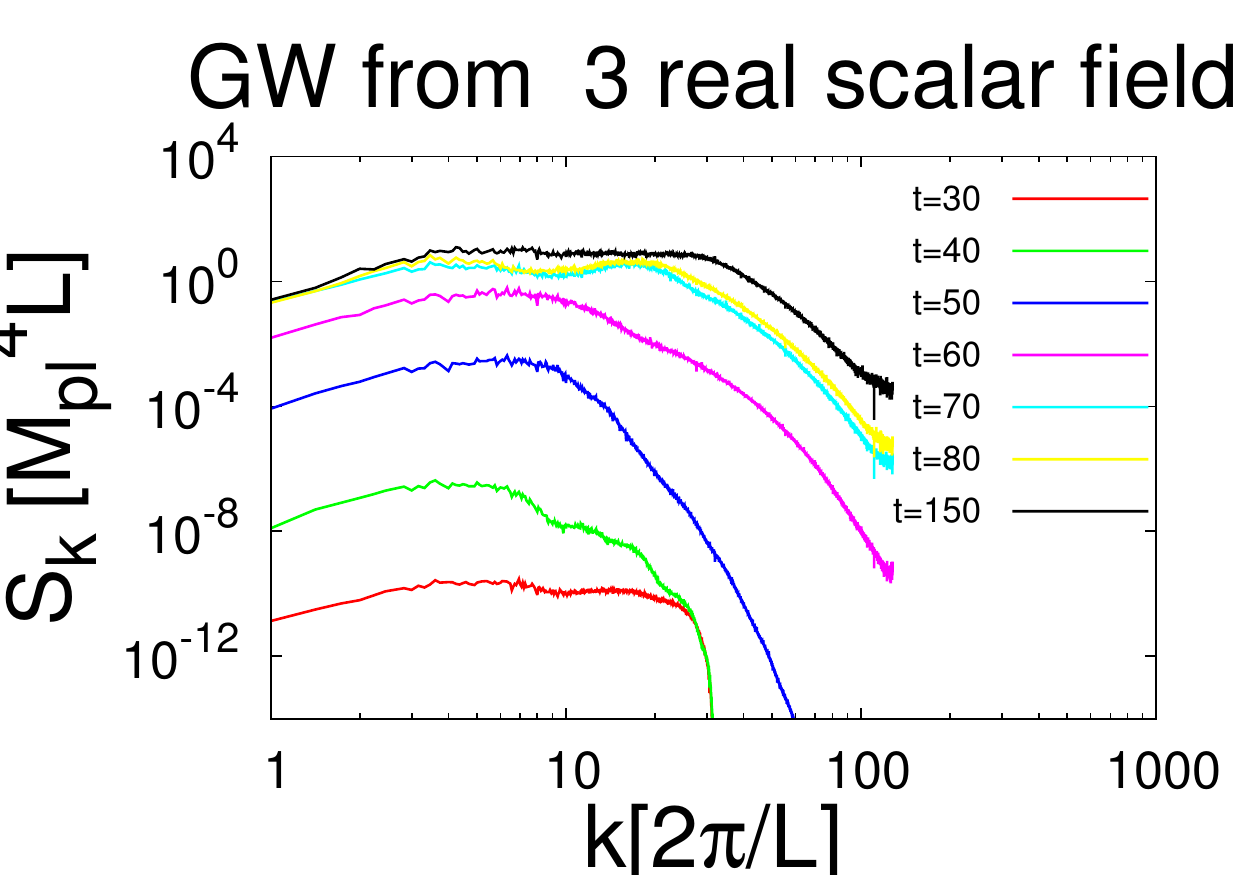}
\caption{
Time evolution of the spectrum of the gravitational waves from 
$t = 30[1/m_{\sigma}]$ to $t=80[1/m_{\sigma}]$ for every $10[1/m_{\sigma}]$ time 
interval and at the final time $t=150[1/m_{\sigma}]$ .
Here, the vertical axis corresponds to $S_k$ defined in (\ref{eq:3:sk}).
}
\label{fig:3:gw}
\end{center}
\end{figure}

Using the result of the lattice simulations, we calculate the spectrum of the 
gravitational waves. 
Fig.~\ref{fig:3:gw} shows the spectrum of the 
gravitational waves produced from three real scalar fields. 
We can see that as the fluctuations of the scalar fields increase after 
the first inflation, gravitational waves are produced abundantly. 
This shows that the enhancement of the scalar fields by the parametric 
resonance works as the origin of the gravitational waves. 
Furthermore, we perform the simulations for 30 realizations with different initial random values of the scalar field fluctuations. 
Using the result of these 30 realizations and (\ref{eq:3:OmegaL}), 
we estimate the amount of the gravitational waves 
$\Omega_{\mathrm{gw}}|_{\mathrm{Lattice}}$ as
\begin{equation}\label{eq:3:Omegap_Lattice1}
\Omega_{\mathrm{gw}}|_{\mathrm{Lattice}}
=
\left(6.1\pm0.1\right)\times10^{-4}.
\end{equation}

With the lattice simulations, we can precisely calculate the energy density and the 
spectrum of the gravitational waves produced during the oscillatory phase. 
However, because of the cost of the simulation time, we cannot perform the lattice 
simulation for all parameters allowed in this model. 
In order to estimate the amount of the gravitational waves for various parameters, 
we use~(\ref{eq:3:Omegap}) determining the numerical constant $\alpha$ from the 
result of the lattice simulations. 
For the parameters given by~(\ref{eq:3:parameter}), 
(\ref{eq:3:Omegap}) leads to 
\begin{equation}\label{eq:3:Omegap_estimate1}
\Omega_{\mathrm{gw}}|_{\mathrm{estimate}}
\simeq
\alpha\times2.6\times10^{-4}.
\end{equation}
Comparing (\ref{eq:3:Omegap_Lattice1}) with (\ref{eq:3:Omegap_estimate1}), 
we can determine the $\alpha$ as
\begin{equation}
\alpha\simeq2.71\pm0.05~.
\end{equation}
Thus, the amount of the gravitational waves just after the production is 
estimated as
\begin{equation}
   \Omega_{\mathrm{gw},p}
   \simeq \left(0.92 \pm 0.02\right)\frac{\mu M}{M_{\mathrm{pl}}^2}.
\end{equation}


\section{Present density  of gravitational waves}
\label{sec:present_density}


After the oscillatory phase, the universe is dominated by the false vacuum energy 
$v^4$ and the new inflation starts. 
After the new inflation, the universe experiences the reheating followed by the 
standard thermal history. 
Because of the cosmic expansion, the gravitational waves are red-shifted and their 
energy density decreases. 
Considering these effects, we calculate the present peak frequency 
of the gravitational wave spectrum $f_{\mathrm{gw},0}$, and the amount of the 
gravitational waves $\Omega_{\mathrm{gw},0}$. 

\subsection{Peak frequency of gravitational waves}

As a result of the lattice simulation, it is found that the spectrum of the 
gravitational waves has a 
sharp peak at $f_{\mathrm{gw},p}\simeq k_{\mathrm{peak}}/2\pi$, 
\begin{equation}
f_{\mathrm{gw},p}
=
\frac{0.35m_{\sigma}}{2\pi}.
\end{equation}
This frequency decreases by the cosmic expansion and the present frequency 
$f_{\mathrm{gw},0}$ is given by $\left(a_{p}/a_0\right)f_{\mathrm{gw},p}$, 
where $a_{p}$ is the scale factor at the production of the gravitational 
waves. 
We assume that the scale factor $a_{p}$ is approximately equal to the 
scale factor at the end of the smooth hybrid inflation $a_{\mathrm{H}}$. 
Then, the ratio $a_0/a_{p}$ is given by 
\begin{equation}
\frac{a_0}{a_{p}}
\simeq
\frac{a_0}{a_{\mathrm{H}}}
=
8.3\times10^{33}~e^{N_{\mathrm{new}}}
\left(\frac{T_{\mathrm{R}}}{10^9\mathrm{GeV}}\right)^{-1/3}
\left(\frac{\mu}{M_{\mathrm{pl}}}\right)^{4/3}.
\end{equation}
Thus, the present peak frequency of the gravitational waves $f_{\mathrm{gw},0}$ is 
given by 
\begin{equation}\label{eq:4:fgw0}
f_{\mathrm{gw},0}
=
6.8\times10^7~e^{-N_{\mathrm{new}}}
\left(\frac{T_{\mathrm{R}}}{10^9\mathrm{GeV}}\right)^{1/3}
\left(\frac{\mu}{M_{\mathrm{pl}}}\right)^{2/3}
\left(\frac{\mu M}{M_{\mathrm{pl}}^2}\right)^{-1/2}\mathrm{Hz}.
\end{equation}

\subsection{Relic density of gravitational waves}

In this subsection, we consider the effect of the cosmic expansion 
on the energy density of the gravitational waves.
Just after the production, the energy density of the gravitational waves decreases 
as  $a^{-4}$, like radiation. 
After the wavelength becomes larger than the horizon during the new 
inflation~(horizon exit), the amplitude of the gravitational waves stops decreasing, 
and the energy density decreases as $a^{-2}$. 
After the end of the new inflation, the gravitational waves re-enter the 
horizon~(horizon entry), and the energy density decreases as $a^{-4}$ again. 
Thus, the energy density of the gravitational waves today $\rho_{\mathrm{gw},0}$ is 
given by
\begin{equation}
\rho_{\mathrm{gw},0} 
=
\left(\frac{a_0}{a_{\mathrm{H}}}\right)^{-4}
\left(\frac{a_{\mathrm{entry}}}{a_{\mathrm{exit}}}\right)^{2}
\rho_{\mathrm{gw},p}.
\end{equation}
Here, $a_{\mathrm{exit}}$ is the scale factor at the horizon exit. 
$a_{\mathrm{entry}}$ is the scale factor at the horizon entry. 

If $N_{\mathrm{new}}$ is sufficiently small, the wavelength never exceeds the horizon during the new inflation. 
We call this case ``No horizon exit". 
As $\sigma_i$ becomes larger, $N_{\mathrm{new}}$ becomes larger and  
horizon exit occurs during the new inflation. 
After the new inflation, the horizon begins to grow, and the gravitational 
waves re-enter the horizon. 
If $N_{\mathrm{new}}$ is small, the horizon size at the end of the new inflation 
is not so 
large compared with the wavelength of the gravitational waves. 
Thus, the gravitational waves re-enter the horizon soon after the new inflation, i.e., before the reheating is completed. 
We call this case ``Reheating time horizon entry". 
As $N_{\mathrm{new}}$ becomes larger, the horizon entry occurs later, i.e. during the radiation dominated era. 
We call this case ``Radiation time horizon entry". 
When $N_{\mathrm{new}}$ becomes much larger, the horizon entry occurs during 
matter dominated era. 
We call this case ``Matter time horizon entry". 
In each case, we can write down the $\Omega_{\mathrm{gw},0}$, as 
\begin{equation}\label{eq:4:Omega0}
\left\{
\begin{array}{ll}
\hspace{-0.5\intextsep}\mathrm{No}~\mathrm{exit} &\hspace{-0.5\intextsep}
h^2\Omega_{\mathrm{gw},0,\mathrm{no}}
=
4.5\times10^{-17}~\mathrm{e}^{-4N_{\mathrm{new}}}
\left(\frac{T_{\mathrm{R}}}{10^9\mathrm{GeV}}\right)^{4/3}
\left(\frac{\mu}{M_{\mathrm{pl}}}\right)^{-4/3}
\Omega_{\mathrm{gw},\mathrm{p}},\\[1.0\intextsep]
&\hspace{-10ex}
	{\rm for}~~
	\mathrm{e}^{N_{\rm new}} <
	1.7 ~ \left(\frac{v}{\mu}\right)^{-2/3}
    \left(\frac{\mu M}{M_{\rm pl}^2}\right)^{-1/2},
    \\[1.0\intextsep]
\hspace{-0.5\intextsep}\mathrm{Reh.}~\mathrm{entry} &\hspace{-0.5\intextsep}
h^2\Omega_{\mathrm{gw},0,\mathrm{reh}}
=
6.4\times10^{-19}~\mathrm{e}^{2N_{\mathrm{new}}}
\left(\frac{v}{\mu}\right)^{4}
\left(\frac{T_{\mathrm{R}}}{10^9\mathrm{GeV}}\right)^{4/3}
\left(\frac{\mu}{M_{\mathrm{pl}}}\right)^{-4/3}
\left(\frac{\mu M}{M_{\mathrm{pl}}^2}\right)^3
\Omega_{\mathrm{gw},\mathrm{p}},\\[1.0\intextsep]
&\hspace{-10ex}{\rm for}~~
	1.7 ~ \left(\frac{v}{\mu}\right)^{-2/3}
    \left(\frac{\mu M}{M_{\rm pl}^2}\right)^{-1/2}
		<
		\mathrm{e}^{N_{\rm new}}
		<
		1.6\times 10^6 \left(
			\frac{T_{\mathrm{R}}}{10^{9}{\rm GeV}}
		\right)^{-2/3}
     \left(\frac{\mu}{M_{\mathrm{pl}}}\right)^{2/3}
     \left(\frac{\mu M}{M_{\rm pl}^2}\right)^{-1/2},
		\\[1.0\intextsep]
\hspace{-0.5\intextsep}\mathrm{Ra.}~\mathrm{entry} &\hspace{-0.5\intextsep}
h^2\Omega_{\mathrm{gw},0,\mathrm{ra}}
=
4.0\times10^{-6}
\left(\frac{v}{\mu}\right)^{4}
\left(\frac{\mu M}{M_{\mathrm{pl}}^2}\right)^2
\Omega_{\mathrm{gw},\mathrm{p}},\\[1.0\intextsep]
&\hspace{-10ex}{\rm for}~~
	1.6\times 10^6 \left(
			\frac{T_{\mathrm{R}}}{10^{9}{\rm GeV}}
		\right)^{-2/3}
     \left(\frac{\mu}{M_{\mathrm{pl}}}\right)^{2/3}
     \left(\frac{\mu M}{M_{\rm pl}^2}\right)^{-1/2}
		<
		\mathrm{e}^{N_{\rm new}}
		\\[.5\intextsep]
		&\hspace{10ex}
		<
			3.2\times 10^{24} ~
      \left(\frac{T_{\rm R}}{10^{9}{\rm GeV}}\right)^{1/3}
      \left(\frac{\mu}{M_{\rm pl}}\right)^{2/3}
      \left(\frac{\mu M}{M_{\rm pl}^2}\right)^{-1/2},
		\\[1.0\intextsep]
\hspace{-0.5\intextsep}\mathrm{Ma.}~\mathrm{entry} &\hspace{-0.5\intextsep}
h^2\Omega_{\mathrm{gw},0,\mathrm{ma}}
=
8.7\times10^{-55}~\mathrm{e}^{2N_{\mathrm{new}}}
\left(\frac{v}{\mu}\right)^{4}
\left(\frac{T_{\mathrm{R}}}{10^9\mathrm{GeV}}\right)^{-2/3}
\left(\frac{\mu}{M_{\mathrm{pl}}}\right)^{-4/3}
\left(\frac{\mu M}{M_{\mathrm{pl}}^2}\right)^3
\Omega_{\mathrm{gw},\mathrm{p}},\\[1.0\intextsep]
&\hspace{-10ex}{\rm for}~~
	3.2\times 10^{24} ~
  \left(\frac{T_{\rm R}}{10^{9}{\rm GeV}}\right)^{1/3}
  \left(\frac{\mu}{M_{\rm pl}}\right)^{2/3}
  \left(\frac{\mu M}{M_{\rm pl}^2}\right)^{-1/2}
	<
	\mathrm{e}^{N_{\rm new}}.
\end{array}
\right.
\end{equation}

\subsection{$f_{\mathrm{gw},0}$ and $\Omega_{\mathrm{gw},0}$}

\begin{figure}[tbp]
\begin{center}
\includegraphics[width=120mm]{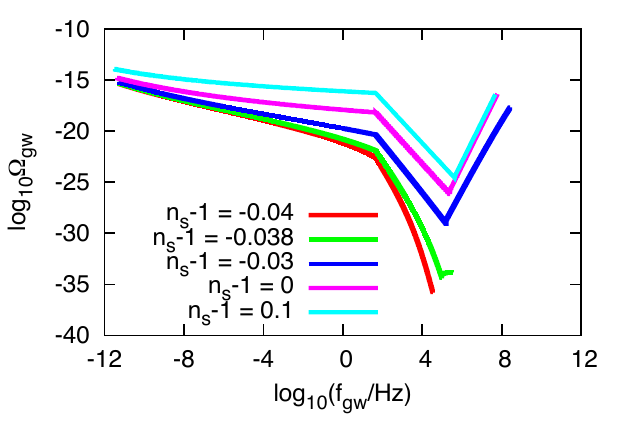}
\caption{
Relations between~$\Omega_{\mathrm{gw},0}$~and~$f_{\mathrm{gw},0}$ for given spectral indices $n_s= -0.04, -0.038, -0.03,  ~0$ and $0.1$.
}
\label{fig:Omega_fg}
\end{center}
\end{figure}

In this subsection, we calculate the relation between the amount of the 
gravitational waves today $\Omega_{\mathrm{gw},0}$ and the peak frequency of the 
gravitational wave spectrum $f_{\mathrm{gw},0}$. 
In~(\ref{eq:4:fgw0}) and~(\ref{eq:4:Omega0}), $f_{\mathrm{gw},0}$ and 
$\Omega_{\mathrm{gw},0}$ are written in terms of 
$N_{\mathrm{new}},~\mu,~v,~M$ and $T_{\mathrm{R}}$. 
Furthermore, in~(\ref{eq:Nn}),~(\ref{eq:mu}) and~(\ref{eq:M}), 
$N_{\mathrm{new}},~\mu,~M$ are written in terms of  $\sigma_{i}$ and $n_s$. 
Here, we assume that the energy scale of the new inflation $v$ is approximately 
equal to that of the hybrid inflation $\mu$, and the reheating temperature 
$T_{\mathrm{R}}$ is $10^9\mathrm{GeV}$. 
For $\mu \simeq v$ the density of the gravitational waves is maximal so that our estimation gives the upper bound. 
With these assumptions, $f_{\mathrm{gw},0}$ and $\Omega_{\mathrm{gw},0}$ are written 
only by $\sigma_{i}$ and $n_s$ as 
\begin{equation}
\ln f_{\mathrm{gw},0}
=
-39.3-\frac{3}{2}\ln\sigma_{i} 
-\frac{1}{4}\ln\left[3\sigma_{i}^2-\left(n_s-1\right)\right]+N_{\mathrm{H}},
\end{equation}
and 
\begin{equation}
\left\{
\begin{array}{ll}
\ln\Omega_{\mathrm{gw},0,\mathrm{no}}
&\hspace{-2ex}=
\ln\alpha -250 +3\ln\sigma_{i} 
+\frac{1}{2}\ln\left[3\sigma_{i}-\left(n_s-1\right)\right]
-2\ln\left[8\sigma_{i}^2-\left(n_s-1\right)\right]
+4N_{\mathrm{H}}\\[0.5\intextsep]
\ln\Omega_{\mathrm{gw},0,\mathrm{reh}}
&\hspace{-2ex}=
\ln\alpha +70.1 +12\ln\sigma_{i} 
+2\ln\left[3\sigma_{i}-\left(n_s-1\right)\right]
-2N_{\mathrm{H}}\\[0.5\intextsep]
\ln\Omega_{\mathrm{gw},0,\mathrm{ra}}
&\hspace{-2ex}=
\ln\alpha -15.12 +9\ln\sigma_{i}
+\frac{3}{2}\ln\left[3\sigma_{i}^2-\left(n_s-1\right)\right]
\\[0.5\intextsep]
\ln\Omega_{\mathrm{gw},0,\mathrm{ma}}
&\hspace{-2ex}=
\ln\alpha -12.5 +12\ln\sigma_{i}
+2\ln\left[3\sigma^2-\left(n_s-1\right)\right] -2N_{\mathrm{H}}
\end{array}
\right.
\end{equation}
where domain of $\sigma_{i}$ is constrained as shown in Fig.~\ref{fig:simami}. 
Thus, for a given spectral index $n_s$ we can obtain  $\Omega_{\mathrm{gw},0}$ as a function of  the peak frequency $f_{\mathrm{gw},0}$. 
Fig.~\ref{fig:Omega_fg} shows this relation. 
At low peak frequencies $f_{{\rm gw},0} \lesssim 40~\rm{Hz}$, the horizon entry occurs during the radiation dominated era. 
At high peak frequencies $40~{\rm Hz} \lesssim f_{{\rm gw},0} \lesssim 10^5~{\rm Hz}$, the horizon entry occurs during the reheating era. 
At higher peak frequencies $f_{{\rm gw},0} \gtrsim 10^5~\rm{Hz}$, the horizon exit does not occur. 
In this model, as $\sigma_{i}$ becomes larger, $\mu$ and $\Omega_{\mathrm{gw},p}$ 
becomes larger. 
In addition to this, as $\sigma_{i}$ becomes larger, 
$N_{\mathrm{new}}$ becomes larger. 
Hence, $f_{\mathrm{gw},0}$ becomes smaller as shown in (\ref{eq:4:fgw0}). 
When the wavelength is larger than the horizon, the amplitude of the 
gravitational waves does not decrease. 
Because of this effect, $\Omega_{\mathrm{gw},p}$ increases. 
Thus, the curves of the plot have negative slope. 
When horizon exit does not occur, the amplitude of the 
gravitational waves always decreases during the new inflation. 
This effect decreases the amount of the gravitational waves more efficiently than in 
the case where horizon exit occurs, and the slope of the plot becomes positive 
at higher frequencies.


\section{Conclusion}
\label{sec:conclusion}

In this paper, we have studied the production of the gravitational waves 
in the smooth hybrid new inflation model \cite{Yamaguchi:2004tn,Kawasaki:2006zv}. 
After the smooth hybrid inflation, the inflaton and waterfall fields for the 
hybrid inflation start to oscillate and their 
fluctuations grow exponentially 
through the parametric resonance, which leads to the efficient production of 
gravitational waves. 
We performed the lattice simulation and calculated the amount of 
the gravitational waves. 

\begin{figure}[tb]
\begin{center}
\includegraphics[angle=-90,width=120mm]{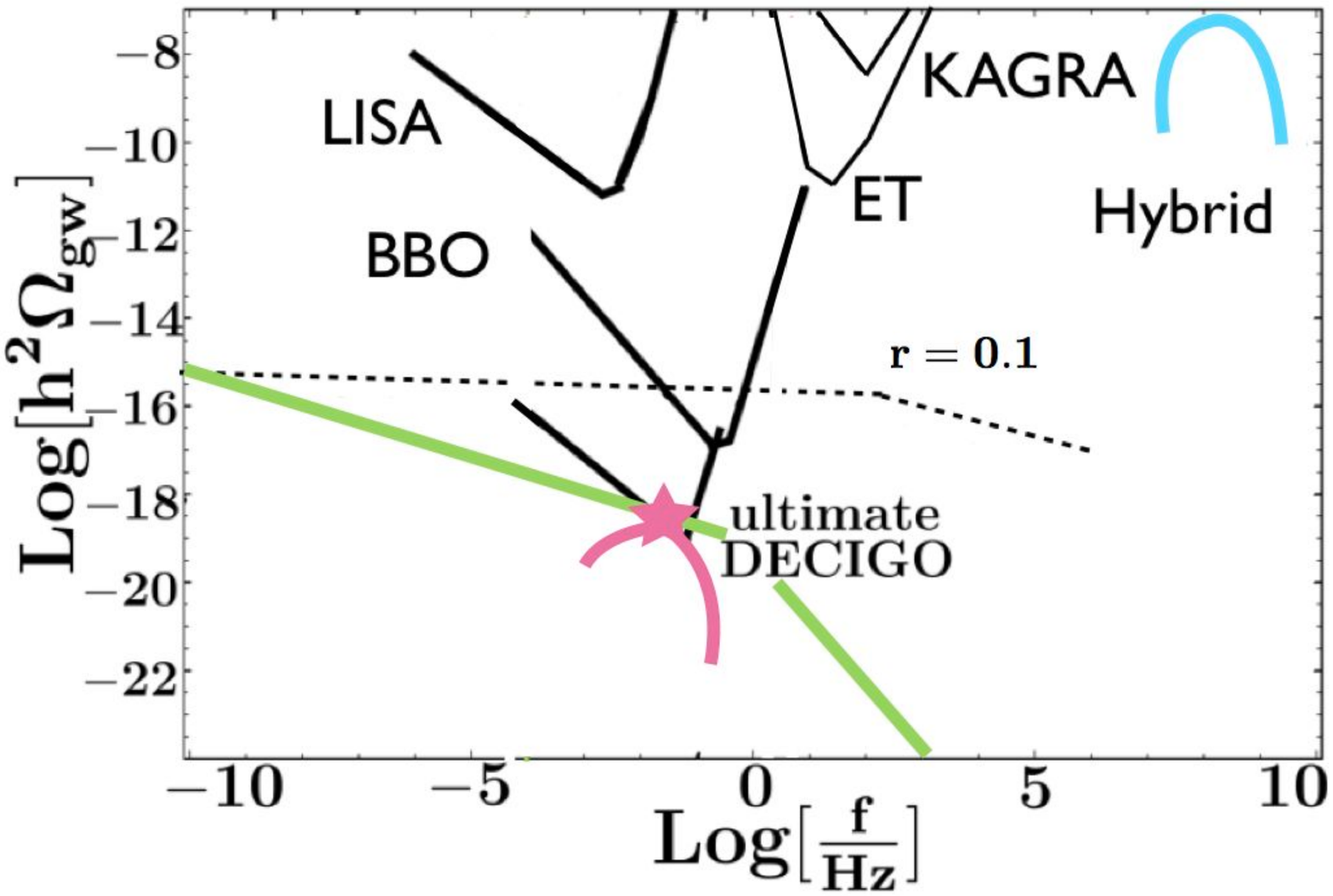}
\end{center}
\caption{Sensitivity curves of planned detectors, ultimate DECIGO, BBO, 
LISA, ET, and KAGRA. 
The green line shows the amplitude of the gravitational waves at the peak 
frequency for each set of the parameters, especially the pink line shows 
the spectra of the gravitational waves in the present model with 
$\mu=3.8\times10^{-4}M_{{\rm pl}},~M=4.1\times10^{-2}M_{{\rm pl}}$ and 
$n_s-1=-0.035$.
The blue line shows the spectrum of the gravitational waves from the 
single hybrid inflation~\cite{Dufaux:2008dn}. 
The dotted line shows the upper bound on the tensor to scalar ratio.
}
\label{fig:sensiv}
\end{figure}

From the result of the lattice simulation, 
we found the relation between 
the present gravitational energy density $\Omega _{gw,0}$ 
and the peak frequency $f_{gw}$. 
Fig.~\ref{fig:sensiv} shows the sensitivities of the planned detectors 
and gravitational wave spectrum predicted from inflation models. 
This figure shows that the gravitational waves produced in this model can be 
detectable in the future experiments in some parameter space. 
For example, the pink line shows the spectrum of gravitational waves 
produced in a set of parameters $\mu = 3.8\times10^{-4}M_{{\rm pl}},~M=4.1\times10^{-2}M_{{\rm pl}}
,~n_s-1=-0.035$, where $n_s$ is the best-fit value of the Planck result. 
In this parameter set, the ultimate DECIGO can detect the gravitational waves.

\section*{Acknowledgments}
This work is supported by Grant-in-Aid for Scientific research from
the Ministry of Education, Science, Sports, and Culture (MEXT), Japan,
No.\ 14102004 (M.K.), No.\ 21111006 (M.K.) and also 
by World Premier International Research Center
Initiative (WPI Initiative), MEXT, Japan. 
K.S. is supported by the Japan Society for the Promotion of Science (JSPS).



\begin{thebibliography}{99}
\bibitem{Hulse:1974eb}
  R.~A.~Hulse and J.~H.~Taylor,
  Astrophys.\ J.\  {\bf 195} (1975) L51.
  
\bibitem{LIGO}
  http://www.ligo.caltech.edu/
  
 \bibitem{Virgo}
  https://wwwcascina.virgo.infn.it/
  
\bibitem{ET}
  http://www.et-gw.eu/

\bibitem{KAGRA}
  http://gwcenter.icrr.u-tokyo.ac.jp/

\bibitem{LISA}
	http://lisa.nasa.gov/
  
\bibitem{DECIGO1}
  http://tamago.mtk.nao.ac.jp/decigo
  
\bibitem{Seto:2001qf}
  N.~Seto, S.~Kawamura and T.~Nakamura,
  Phys.\ Rev.\ Lett.\  {\bf 87} (2001) 221103
  [astro-ph/0108011].
  
\bibitem{Kawamura:2011zz}
  S.~Kawamura, M.~Ando, N.~Seto, S.~Sato, T.~Nakamura, K.~Tsubono, N.~Kanda and T.~Tanaka {\it et al.},
  Class.\ Quant.\ Grav.\  {\bf 28} (2011) 094011.
  
\bibitem{Crowder:2005nr}
  J.~Crowder and N.~J.~Cornish,
  Phys.\ Rev.\ D {\bf 72} (2005) 083005
  [gr-qc/0506015].
  
\bibitem{Ade:2013zuv}
  P.~A.~R.~Ade {\it et al.}  [Planck Collaboration],
		  arXiv:1303.5076 [astro-ph.CO].

\bibitem{Lyth:2009zz}
  D.~H.~Lyth and A.~R.~Liddle,
  Cambridge, UK: Cambridge Univ. Pr. (2009) 497 p  
  
\bibitem{Khlebnikov:1997di}
  S.~Y.~Khlebnikov and I.~I.~Tkachev,
		  Phys.\ Rev.\ D {\bf 56} (1997) 653
			  [hep-ph/9701423].

\bibitem{Easther:2006gt}
  R.~Easther and E.~A.~Lim,
		  JCAP {\bf 0604} (2006) 010
			  [astro-ph/0601617].

\bibitem{GarciaBellido:2007af}
  J.~Garcia-Bellido, D.~G.~Figueroa and A.~Sastre,
		  Phys.\ Rev.\ D {\bf 77} (2008) 043517
			  [arXiv:0707.0839 [hep-ph]].

\bibitem{Dufaux:2008dn}
  J.~-F.~Dufaux, G.~Felder, L.~Kofman and O.~Navros,
  JCAP {\bf 0903} (2009) 001
  [arXiv:0812.2917 [astro-ph]].
  
\bibitem{Hiramatsu:2010yz}
  T.~Hiramatsu, M.~Kawasaki and K.~'i.~Saikawa,
  JCAP {\bf 1005} (2010) 032
  [arXiv:1002.1555 [astro-ph.CO]].

\bibitem{Kawasaki:2011vv}
  M.~Kawasaki and K.~'i.~Saikawa,
  JCAP {\bf 1109} (2011) 008
  [arXiv:1102.5628 [astro-ph.CO]].
  
\bibitem{Berezinsky:2000vn}
  V.~Berezinsky, B.~Hnatyk and A.~Vilenkin,
		  astro-ph/0001213.

\bibitem{Damour:2000wa}
  T.~Damour and A.~Vilenkin,
		  Phys.\ Rev.\ Lett.\  {\bf 85} (2000) 3761
			  [gr-qc/0004075].
\bibitem{Damour:2004kw}
  T.~Damour and A.~Vilenkin,
		  Phys.\ Rev.\ D {\bf 71} (2005) 063510
			  [hep-th/0410222].

\bibitem{Kawasaki:2010yi}
  M.~Kawasaki, K.~Miyamoto and K.~Nakayama,
  Phys.\ Rev.\ D {\bf 81} (2010) 103523
  [arXiv:1002.0652 [astro-ph.CO]].
  
\bibitem{GarciaBellido:1997wm}
  J.~Garcia-Bellido and A.~D.~Linde,
		  Phys.\ Rev.\ D {\bf 57} (1998) 6075
			  [hep-ph/9711360].

\bibitem{Felder:2000hj}
  G.~N.~Felder, J.~Garcia-Bellido, P.~B.~Greene, L.~Kofman, A.~D.~Linde and I.~Tkachev,
		  Phys.\ Rev.\ Lett.\  {\bf 87} (2001) 011601
			  [hep-ph/0012142].
						
\bibitem{Randall:1995dj}
  L.~Randall, M.~Soljacic and A.~H.~Guth,
		  Nucl.\ Phys.\ B {\bf 472} (1996) 377
			  [hep-ph/9512439].
						
\bibitem{GarciaBellido:1996qt}
  J.~Garcia-Bellido, A.~D.~Linde and D.~Wands,
		  Phys.\ Rev.\ D {\bf 54} (1996) 6040
			  [astro-ph/9605094].

\bibitem{Izawa:1997df}
  K.~I.~Izawa, M.~Kawasaki and T.~Yanagida,
  Phys.\ Lett.\ B {\bf 411} (1997) 249
  [hep-ph/9707201].
  
\bibitem{Yamaguchi:2004tn}
  M.~Yamaguchi and J.~'i.~Yokoyama,
  Phys.\ Rev.\ D {\bf 70} (2004) 023513
  [hep-ph/0402282].

\bibitem{Lazarides:1995vr}
  G.~Lazarides and C.~Panagiotakopoulos,
		  Phys.\ Rev.\ D {\bf 52} (1995) 559
			  [hep-ph/9506325].

\bibitem{Jeannerot:2000sv}
  R.~Jeannerot, S.~Khalil, G.~Lazarides and Q.~Shafi,
		  JHEP {\bf 0010} (2000) 012
			  [hep-ph/0002151].

\bibitem{Kawasaki:2006zv}
  M.~Kawasaki, T.~Takayama, M.~Yamaguchi and J.~'i.~Yokoyama,
  Phys.\ Rev.\ D {\bf 74} (2006) 043525
  [hep-ph/0605271].

\bibitem{Linde:1997sj}
  A.~D.~Linde and A.~Riotto,
		  Phys.\ Rev.\ D {\bf 56} (1997) 1841
			  [hep-ph/9703209].


\bibitem{Izawa:1996dv}
  K.~-I.~Izawa and T.~Yanagida,
  Phys.\ Lett.\ B {\bf 393} (1997) 331
  [hep-ph/9608359].

\bibitem{Landau}
  L.Landau and E.Lifschitz, Mechanics (Pergamon, Oxford, 1960);

\bibitem{Shtanov:1994ce}
  Y.~Shtanov, J.~H.~Traschen and R.~H.~Brandenberger,
  Phys.\ Rev.\ D {\bf 51} (1995) 5438
  [hep-ph/9407247].

\bibitem{Kofman:1997yn}
  L.~Kofman, A.~D.~Linde and A.~A.~Starobinsky,
  Phys.\ Rev.\ D {\bf 56} (1997) 3258
  [hep-ph/9704452].
  
\bibitem{Felder:2006cc}
  G.~N.~Felder and L.~Kofman,
  Phys.\ Rev.\ D {\bf 75} (2007) 043518
  [hep-ph/0606256].
  
\bibitem{Hiramatsu:2010yn}
  T.~Hiramatsu, M.~Kawasaki and K.~'i.~Saikawa,
  JCAP {\bf 1108} (2011) 030
  [arXiv:1012.4558 [astro-ph.CO]].

\bibitem{Dufaux:2007pt}
  J.~F.~Dufaux, A.~Bergman, G.~N.~Felder, L.~Kofman and J.~-P.~Uzan,
  Phys.\ Rev.\ D {\bf 76} (2007) 123517
  [arXiv:0707.0875 [astro-ph]].
  
\bibitem{Polarski:1995jg}
  D.~Polarski and A.~A.~Starobinsky,
		  Class.\ Quant.\ Grav.\  {\bf 13} (1996) 377
			  [gr-qc/9504030].

\bibitem{Khlebnikov:1996mc}
  S.~Y.~.Khlebnikov and I.~I.~Tkachev,
		  Phys.\ Rev.\ Lett.\  {\bf 77} (1996) 219
			  [hep-ph/9603378].

\bibitem{Khlebnikov:1996zt}
  S.~Y.~.Khlebnikov and I.~I.~Tkachev,
		  Phys.\ Rev.\ Lett.\  {\bf 79} (1997) 1607
			  [hep-ph/9610477].

\bibitem{GarciaBellido:2002aj}
  J.~Garcia-Bellido, M.~Garcia Perez and A.~Gonzalez-Arroyo,
		  Phys.\ Rev.\ D {\bf 67} (2003) 103501
			  [hep-ph/0208228].
  
\end{thebibliography}
\end{document}